\documentclass[preprint,proceedings]{rmaa}

\suppressfulladdresses 

\usepackage{paralist}

\usepackage{psfrag,color}

\usepackage{natbib,makeidx}
\usepackage{lscape}
\usepackage{epsfig}
\usepackage[rightcaption]{sidecap}

\SetYear{2006}
\SetConfTitle{Serie de Conferencias}

\title{Old Main-Sequence Turnoff Photometry in the SMC} 

\author{
  N. E. D. No\"el,\altaffilmark{1} 
  C. Gallart,\altaffilmark{1}
  E. Costa,\altaffilmark{2}
  R. A. M\'endez\altaffilmark{2}}

\altaffiltext{1}{Instituto de Astrof\'\i{}sica de Canarias,
  38200, La Laguna, Tenerife, Spain (noelia@iac.es).}

\altaffiltext{2}{Universidad
de Chile, Casilla 36-D, Las Condes, Santiago, Chile.}

\shortauthor{No\"el, Gallart, Costa, \& M\'endez}
\shorttitle{RevMexAA(SC) Demo Document}

\listofauthors{N. E. D. No\"el, C. Gallart, E. Costa, \& R. A. M\'endez}
\indexauthor{No\"el, N. E. D.}
\indexauthor{Gallart, C.}
\indexauthor{Costa, E.}
\indexauthor{M\'endez, R. A.}

\abstract{We present ground-based {\it B} and {\it R}-band color-magnitude diagrams (CMDs) of unprecedented depth 
for twelve fields in the Small Magellanic Cloud (SMC). They 
 reach the oldest main-sequence turnoffs and cover
 a wide range of galactocentric distances
up to $\sim$4$\arcdeg$ from the SMC center, and are located at different position angles. 
A picture of the stellar content in our SMC fields is presented, through the comparison with theoretical
isochrones. 
Our study confirms the existence of strong population
gradients and spatial variation in the SMC stellar content. 
 None of the SMC fields presented here are dominated by
 old stellar populations which proves that at $\sim$4$\arcdeg$ from the SMC center we do not reach an old stellar
  halo similar to that of the Milky Way.}

\resumen{Se presenta un conjunto de diagramas color-magnitud (DCM)  
correspondientes a doce campos localizados en 
 la Peque\~na Nube de Magallanes (SMC). Estos DCM poseen
  una profundidad sin precedentes,
  llegando hasta 
los puntos de giro m\'as viejos de la secuencia principal y cubriendo un amplio
rango de distancias galactoc\'entricas,
hasta $\sim$4$\arcdeg$ desde el centro de la SMC y con diferentes \'angulos de posici\'on.
Se expone un an\'alisis del contenido estelar en estos campos de la SMC a trav\'es de la comparaci\'on con isocronas te\'oricas.
 Nuestro estudio confirma la existencia de fuertes gradientes de poblaci\'on y variaciones espaciales en el contenido estelar
de la SMC. 
Ninguno de los campos mostrados est\'a
dominado por una poblaci\'on estelar exclusivamente vieja. Esto prueba que a $\sim$4$\arcdeg$ del centro de la SMC no hemos alcanzado un
halo estelar viejo similar al de la V\'\i{}a L\'actea.}

\addkeyword{Local group galaxies: evolution}
\addkeyword{galaxies: individual (SMC)}
\addkeyword{galaxies: stellar content}

\begin{document}
\maketitle

\section{Introduction}
\label{sec:intro}

Besides the intrinsec interest in studying
 the SMC stellar populations, which is key to understand our local environment, this galaxy has been largely
 neglected in favour of its closest neighbour, the LMC. 
There is a disagreement in the published SFHs of the SMC since some authors (e.g. Harris \& Zaritsky 2004) argue that 
there was a quiescent period at intermediate ages in which the galaxy formed few or no stars,
 and others (e.g. Dolphin et al. 2001) do not find a gap in the star formation rate of the SMC. It is important to stress that the SMC 
 regions analised by these authors are different. 
In an attempt to address this conflict, here we present a discussion based on deep CMDs with theoretical
 isochrones overlapped.

\section{SMC Stellar Content}
\label{sec:stellar}

\begin{SCfigure*}[1][t]

\begin{wide}
\psfig{figure= 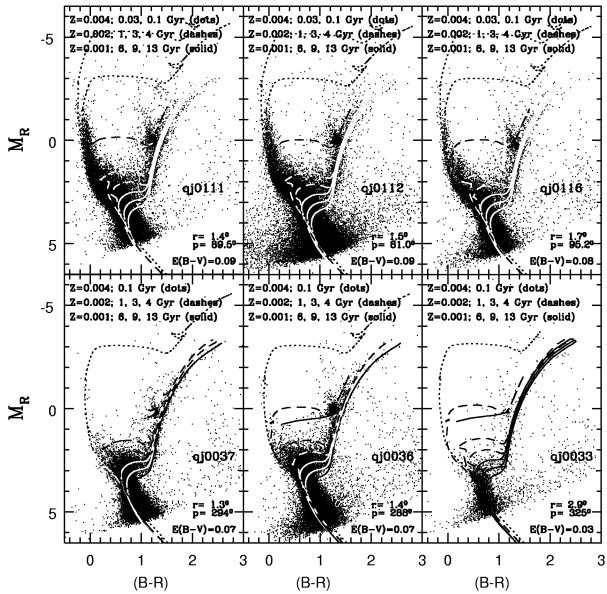, width=0.85\linewidth}
  \caption[Short line]{Upper panel shows the CMDs of the Eastern SMC fields, located 
  in a range of $\sim$1.4$\arcdeg$ to $\sim$1.7$\arcdeg$ from the center.
   Canonical isochrones of 6, 9, and 13 Gyr with Z=0.001,
   isochrones with overshooting of 1, 3, and 4 Gyr with Z=0.002,
 and of 0.1 Gyr and 0.03 Gyr with Z=0.004 were overlapped.
 In these CMDs the conspicuous MS is well populated from the oldest turnoff
at M$_{R}$=3.5 up to the 0.03 Gyr isochrone. All the CMDs show
 a large fraction of young stars ($<$1 Gyr old). The densely populated areas around the 
 isochrones and the extension in luminosity of the Red Clump indicate a strong presence of 
intermediate-age stars. 
 Lower panel shows the CMDs of the Western fields, located up to $\sim$2.9$\arcdeg$ from the SMC center.
 In these fields, 
the intermediate-age population ($\sim$3 to 9 Gyr) is dominant.
 Fields qj0036 and qj0037 have a small fraction of stars born up to 0.1 Gyr ago, 
while qj0033, the most remote of the three, seems to have no stars 
younger than 1 Gyr and the main population is composed by stars between 6 and 9 Gyr old.
 Position angles (p), reddening (E(B-R)), metallicities and ages of the
 isochrones are labeled. A distance modulus (m-M$_{0}$)=18.9 has been assumed. \label{fig:fig1}}

\end{wide}
\end{SCfigure*}

\begin{SCfigure*}[1][t]

\begin{wide}
\psfig{figure= 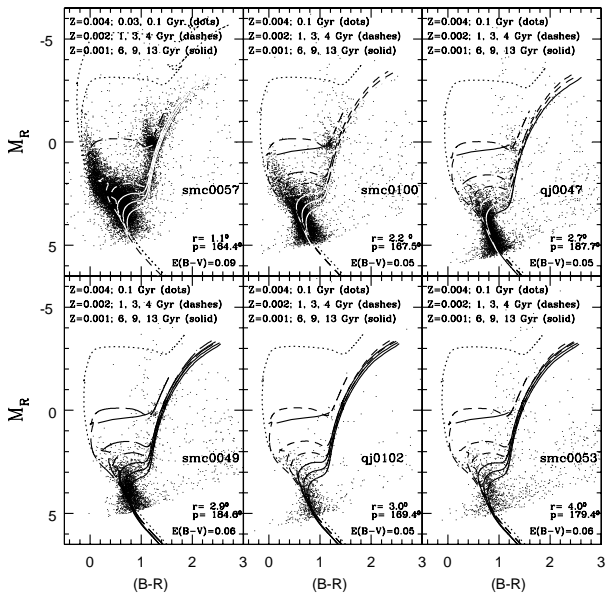, width=0.85\linewidth}
  \caption[Short line]{Figure showing the CMDs of the six Southern SMC fields. 
   The same isochrones as in Figure~\ref{fig:fig1} have been overlapped.
    In field smc0057 a 0.03 Gyr isochrone with overshooting (Z=0.004) was also superimposed.
      This field is the closest to the SMC center and the one with the smallest position angle (in 
this Southern part), combination that
gives a CMD with an important MS, where the 0.1 Gyr isochrone is still quite populated. 
The population gradient is depicted by the gradual lack of stars from 0.01 to 9 Gyr 
 as going further South from the SMC center.
 Despite this, 
 even in the more distant fields, the population is not purely old but dominated by 
intermediate-age stars as is shown in the zones around the 3, 4, 6, and 9 Gyr isochrones.
Neither of the SMC CMDs in Figure~\ref{fig:fig1} nor in this Figure 
 present a blue horizontal branch. 
 CMDs in both Figures are shown in order of increasing distance from the SMC 
 center. 
 Position angles (p), reddening (E(B-R)), metallicities and ages of the isochrones are labeled.
  A distance modulus (m-M$_{0}$)=18.9 has been assumed. \label{fig:fig2}}
\end{wide}
\end{SCfigure*}

The interpretation of composite stellar population CMDs strongly relies on the stellar evolution models adopted (see
Gallart et al. 2005). For our purpose we used the Teramo stellar evolution models (Pietrinferni et al. 2004) as shown 
 in Figure~\ref{fig:fig1} (Eastern SMC fields in the upper panel and Western SMC fields in the lower panel) and in
  Figure~\ref{fig:fig2} (Southern 
 SMC fields), using metallicities suitable
for the SMC stellar populations. 

Our analysis shows the presence of spatial variations in the stellar content as a function of the position angle and 
strong gradients in the stellar population as a function of the galactocentric distance. 
 The fact that the SMC CMDs analyzed here do not show a blue horizontal branch and that none of them is
 dominated by a completely old population indicate that,
at $\sim$4$\arcdeg$ from the SMC center, we do not reach an old halo similar to that of the Milky Way (No\"el et al. 2006).

\end{document}